\documentclass[a4paper,12pt, epsfig]{article}
\usepackage{epsfig}
\usepackage{amssymb}
\usepackage{amsfonts}
\usepackage{amsmath}
\usepackage{euscript}
\usepackage{verbatim}
\usepackage{latexsym}
\usepackage{graphicx}

\newskip\humongous \humongous=0pt plus 1000pt minus 1000pt

\newif\ifdtup

\catcode`\@=11

\@addtoreset{equation}{section}

\def\@normalsize{\@setsize\normalsize{15pt}\xiipt\@xiipt
\abovedisplayskip 14pt plus3pt minus3pt%
\belowdisplayskip \abovedisplayskip
\abovedisplayshortskip \z@ plus3pt%
\belowdisplayshortskip 7pt plus3.5pt minus0pt}

\def\small{\@setsize\small{13.6pt}\xipt\@xipt
\abovedisplayskip 13pt plus3pt minus3pt%
\belowdisplayskip \abovedisplayskip
\abovedisplayshortskip \z@ plus3pt%
\belowdisplayshortskip 7pt plus3.5pt minus0pt
\def\@listi{\parsep 4.5pt plus 2pt minus 1pt
     \itemsep \parsep
     \topsep 9pt plus 3pt minus 3pt}}

\relax

\catcode`@=12

\catcode`\@=11

\def\section{\@startsection{section}{1}{\z@}{3.5ex plus 1ex minus
   .2ex}{2.3ex plus .2ex}{\large\bf}}

\def\SymBoxes#1#2#3#4{\newdimen\un@t \un@t#3%
\raisebox{#1}{\rule{#2\un@t}{#4}\hskip-#2\un@t
\@tempdimb\un@t \advance\@tempdimb by-#4\@tempcntb#2\relax%
\@whilenum{\@tempcntb>0}\do{
\rule{#4}{\un@t}\hskip\@tempdimb \advance\@tempcntb by\m@ne}%
\hskip-#2\un@t \rule[\un@t]{#2\un@t}{#4}%
\rule[\un@t]{#4}{#4}\hskip-#4
\rule{#4}{\un@t}}\hskip-#4}                

\begin{document}

\newcommand{\beq}{\begin{equation}}
\newcommand{\eeq}{\end{equation}}
\newcommand{\bea}{\begin{eqnarray}}
\newcommand{\eea}{\end{eqnarray}}
\newcommand{\beas}{\begin{eqnarray*}}
\newcommand{\eeas}{\end{eqnarray*}}
\newcommand{\defi}{\stackrel{\rm def}{=}}
\newcommand{\non}{\nonumber}
\newcommand{\bquo}{\begin{quote}}
\newcommand{\enqu}{\end{quote}}
\renewcommand{\(}{\begin{equation}}
\renewcommand{\)}{\end{equation}}
\def \eqn#1#2{\begin{equation}#2\label{#1}\end{equation}}
\def\IZ{{\mathbb Z}}
\def\IR{{\mathbb R}}
\def\IC{{\mathbb C}}
\def\IQ{{\mathbb Q}}
\def\de{\partial}
\def\Tr{ \hbox{\rm Tr}}
\def\H{ \hbox{\rm H}}
\def\HE{ \hbox{$\rm H^{even}$}}
\def\HO{ \hbox{$\rm H^{odd}$}}
\def\K{ \hbox{\rm K}}
\def\Im{ \hbox{\rm Im}}
\def\Ker{ \hbox{\rm Ker}}
\def\const{\hbox {\rm const.}}
\def\o{\over}
\def\im{\hbox{\rm Im}}
\def\re{\hbox{\rm Re}}
\def\bra{\langle}\def\ket{\rangle}
\def\Arg{\hbox {\rm Arg}}
\def\Re{\hbox {\rm Re}}
\def\Im{\hbox {\rm Im}}
\def\exo{\hbox {\rm exp}}
\def\diag{\hbox{\rm diag}}
\def\longvert{{\rule[-2mm]{0.1mm}{7mm}}\,}
\def\a{\alpha}
\def\dag{{}^{\dagger}}
\def\tq{{\widetilde q}}
\def\p{{}^{\prime}}
\def\W{W}
\def\N{{\cal N}}
\def\hsp{,\hspace{.7cm}}

\def\br{\nonumber\\}
\def\IZ{{\mathbb Z}}
\def\IR{{\mathbb R}}
\def\IC{{\mathbb C}}
\def\IQ{{\mathbb Q}}
\def\IP{{\mathbb P}}
\def \eqn#1#2{\begin{equation}#2\label{#1}\end{equation}}

\newcommand{\C}{\ensuremath{\mathbb C}}
\newcommand{\Z}{\ensuremath{\mathbb Z}}
\newcommand{\R}{\ensuremath{\mathbb R}}
\newcommand{\rp}{\ensuremath{\mathbb {RP}}}
\newcommand{\cp}{\ensuremath{\mathbb {CP}}}
\newcommand{\vac}{\ensuremath{|0\rangle}}
\newcommand{\vact}{\ensuremath{|00\rangle}                    }
\newcommand{\oc}{\ensuremath{\overline{c}}}
\begin{titlepage}
\begin{flushright}
SISSA 26/2011/EP
\end{flushright}
\bigskip
\def\thefootnote{\fnsymbol{footnote}}

\begin{center}
{\Large
{\bf
Quantum Critical Superfluid Flows \\
\vspace{0.2in}
and Anisotropic Domain Walls
}
}
\end{center}

\bigskip
\begin{center}
{\large  Daniel Are\'an$^{a}$,
Matteo Bertolini$^{a,b}$,
Chethan Krishnan$^b$\\
\vspace{0.1in}
and Tom\'{a}\v{s} Proch\'{a}zka$^b$}

\end{center}

\renewcommand{\thefootnote}{\arabic{footnote}}

\begin{center}
$^a$ {International Centre for Theoretical Physics (ICTP)\\
Strada Costiera 11; I 34014 Trieste, Italy \\}
\vskip 5pt
$^b$ {SISSA and INFN - Sezione di Trieste\\
Via Bonomea 265; I 34136 Trieste, Italy\\}
\vskip 5pt
{\texttt{arean,bertmat,krishnan,procht @sissa.it}}

\end{center}

\noindent
\begin{center} {\bf Abstract} \end{center}
We construct charged anisotropic AdS domain walls as solutions of a consistent truncation of type IIB string theory. 
These are a one-parameter family of solutions that flow to an AdS fixed point in the IR, 
exhibiting emergent conformal invariance and quantum criticality. They represent the zero-temperature 
limit of the holographic superfluids at finite superfluid velocity constructed in arXiv:1010.5777. We show that these domain walls exist 
only for velocities less than a critical value, agreeing in detail with a conjecture made there. We also comment about the IR limits of 
flows with velocities higher than this critical value, and point out an intriguing similarity between the phase diagrams of holographic 
superfluid flows and those of ordinary superconductors with imbalanced chemical potential.  
\vspace{1.6 cm}
\vfill

\end{titlepage}

\setcounter{footnote}{0}

\section{Introduction}
\label{intro}

Quantum critical points are expected to be of significance in understanding the ground states of high-$T_c$ superconductors. 
Holographic constructions\footnote{Holographic superconductors were first constructed in \cite{HHH1}.} of such quantum critical 
(hence zero-temperature) superconductors give rise to domain wall solutions, which capture the holographic RG flow from a symmetric 
state in the UV to a symmetry-breaking vacuum in the IR \cite{GubRoch2}. 

Examples of such solutions were presented in \cite{GubRoch,Gauntlett:2009bh}. They arise 
in consistent truncations of type IIB and M-theory containing an abelian gauge field and a charged scalar with a 
symmetry-breaking potential. To trigger the RG flow, the UV theory was deformed by a chemical potential $\mu$. 
That is, the $U(1)$ symmetry is not broken explicitly, rather the theory is deformed by means of an  
uncharged operator $\mu J^0$ (i.e.,  $A_\mu^B J^\mu$ with only  the time component turned on\footnote{The 
superscript $B$ denotes the fact that these are boundary quantities.}). This induces a non-vanishing VEV 
for the charged scalar, hence breaking the $U(1)$ symmetry spontaneously.  The domain walls interpolate 
between two AdS minima, in the UV and IR, and represent the ground states of holographic type IIB and 
M-theory superconductors constructed in \cite{Gubser} 
and \cite{Gaunt}. 

A  natural generalization of this idea is to break the isotropy by turning on a spatial component of the boundary 
vector potential. The corresponding 
solutions can be interpreted 
as the zero temperature limits of holographic superfluid flows (i.e. a superfluid with finite superfluid velocity $\xi$) 
discussed first 
in \cite{Basu, Herzog} and generalized in various ways in \cite{Daniel, Tisza, Daniel2}. In \cite{ABKP}, fully backreacted 
holographic superfluid flow solutions were 
constructed at finite temperature within the same type IIB consistent truncation of \cite{GubRoch}. The advantage of a backreacted solution was that one could systematically lower the temperature in a fully controlled way. Using this approach, strong indications were obtained in \cite{ABKP} that in the zero-temperature limit the IR AdS space found by \cite{GubRoch} is actually robust even when a superfluid velocity is turned on, at least up to a critical value $\xi_c$. Various pieces of evidence were presented that there should exist an anisotropic domain wall  solution interpolating between two AdS spaces, for velocities  in the range $0< \xi < \xi_c$. On the other hand, the holographic superfluid 
flows constructed in \cite{ABKP} exist in the bigger range $0< \xi < \xi_*$, with $\xi_*  > \xi_c$ being the velocity above which superfluidity is destroyed and the system is always in the normal phase. The low temperature behavior of the solutions in the range $\xi_c < \xi < \xi_*$ indicated that a different non AdS-like IR phase emerges at zero temperature. In this range the perturbation induced by $\xi$ turns out to be too strong to be washed out by the RG-flow, so that the anisotropy still survives at small radii,  suggesting the non-existence of any obvious quantum critical point. 

In the present paper we put the speculations of \cite{ABKP} on a concrete footing by explicitly constructing gravitational solutions describing  
the ground state of holographic superfluid flows. We find that anisotropic AdS-to-AdS domain wall solutions indeed exist only in the expected range, $0<\xi<\xi_c$, 
and make various quantitative checks that these are the zero-temperature limits of the superfluid flows 
constructed in \cite{ABKP}. On the contrary, for higher velocities we do not see the emergence of an AdS-like geometry in 
the IR. It is an interesting question to ask whether there is a meaningful IR 
limit in this regime, and if so, whether full solutions can be constructed explicitly. While we did not succeed in finding any meaningful geometry 
emerging in the IR for $\xi_c < \xi< \xi_*$, we offer a few preliminary comments in the concluding section. 
\begin{figure} [ht]
\begin{center}
\includegraphics[height=0.23\textheight]{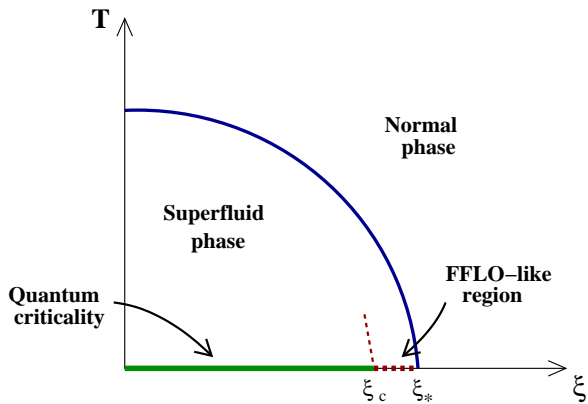}\label{dxi}
\caption{The (qualitative) phase diagram of holographic superfluids. At zero temperature, a quantum critical point is found for velocities below a critical value, $\xi_c$. Above $\xi_c$ 
the system enters a more anisotropic phase where the deformation induced by $\xi$ affects the RG-flow strongly, and brings the system away from any obvious  
AdS-like IR fixed point.}
\end{center}
\end{figure}

Intriguingly, the phase diagram which emerges from our analysis, Figure \ref{dxi}, is reminiscent of 
those expected for real-life superconductors when one induces (e.g., by an external magnetic field) an imbalance $\delta\mu$ 
in the chemical potential between the  
two populations of fermionic particles that form 
Cooper pairs\footnote{We thank A. Cotrone for bringing our attention to these systems.}. For high enough 
$\delta\mu$, still lower than the critical value above which superconductivity is completely destroyed, the system enjoys an anisotropic phase at $T=0$ known as FFLO phase. This phase is characterized by a spontaneous breaking of translational invariance produced by the imbalance $\delta \mu$, and  is separated from the 
ordinary superconducting phase by a  Chandrasekhar-Clogston (CC) bound. In this regime, the system finds it energetically favorable to be in a superconducting anisotropic configuration, where Cooper pairs have a non zero net momentum \cite{comb}.  While this has been proven for BCS superconductors \cite{FFLO}, it is argued to be a generic phenomenon also holding e.g. in high-$T_c$ superconductors.  Although the physical setups are different, the response of this system to $\delta\mu$ is very similar to the response of our holographic setups to a non-vanishing superfluid velocity $\xi$. It would be interesting to investigate this analogy further, and more generally, use holography to 
study superconductors with an imbalanced chemical potential, including the prediction for a CC bound. 

The rest of this paper is organized as follows. In section 2 we present the effective five-dimensional bulk action, our ansatz for the 
various fields, and discuss the IR and UV boundary conditions our solutions should satisfy. Section 3 contains our one-parameter family of 
solutions and a discussion of their properties. Finally in section 4, we  comment on our results and  discuss the regime for 
which AdS domain wall solutions do not seem to exist.

\section{The model: construction and strategy}

The theory we will use is the consistent truncation of low energy type IIB string theory considered in \cite{Gubser}. The action takes the form
\bea
\label{IIBac}
S_{IIB}&=&\int d^5x \sqrt{-g}\bigg[R-{L^2 \over 3}F_{ab}F^{ab}+ {1
\over 4}\left({2L \over 3}\right)^3 \epsilon^{abcde}F_{ab}F_{cd}A_e
+\nonumber
\\
&&- \frac{1}{2}\left((\partial_a \psi)^2 + \sinh^2 \psi(\partial_a
\theta -2 A_a)^2-{6 \over L^2}\cosh^2\left({\psi \over
2}\right)(5-\cosh \psi)\right)\bigg]\,.  \eea
We have set $16 \pi G=1$ and  the conventions are such that $\epsilon^{01234}=1/\sqrt{-g}$, and we have written the charged 
(complex) scalar by splitting the phase and the modulus in the form
$\psi e^{i\theta}$. The Abelian gauge field $A$ is dual to an $R$-symmetry in the boundary field
theory \cite{Gubser} and the scalar field has $R$-charge $R=2$. The general equations of motion that follow from this action (working in the gauge $\theta=0$) can be found in \cite{ABKP}.

In \cite{ABKP}, strong evidence was presented for the existence of an anisotropic AdS domain wall that was a solution of the above Lagrangian. 
Our aim here is to construct it explicitly. To do this, we will write down an ansatz for the metric, gauge field and scalar with only radial 
dependence, so that the resulting equations of motion are ordinary differential equations. We will take the same ansatz as was used in 
\cite{ABKP}. This is equivalent to an anisotropic generalization of the structure considered by Gubser et al. in \cite{GubRoch}. 
Specifically, we take
\beq
ds^2=-{r^2 f(r) \over L^2}dt^2+{L^2 h(r)^2 \over r^2 f(r)} dr^2 -2
C(r) \frac{r^2}{L^2}dt\,dx+{r^2 \over L^2}B(r) dx^2+{r^2 \over L^2}
dy^2 + {r^2 \over L^2} dz^2
\label{ModTisza4}
\eeq
for the metric, and 
\beq
A=A_t(r)\, dt + A_x(r) \,dx\;, \qquad \psi=\psi(r)\,,
\label{gauge3}
\eeq
for the gauge and the scalar fields.

With this ansatz, the full set of equations of motion can be massaged to the form of two first order and five second order differential equations for seven functions. Since the structure of the ansatz is the same, the equations of motion we get are also the same as in \cite{ABKP}. As discussed there, these equations exhibit four scaling symmetries and we quote them here for convenience
\bea && t \rightarrow
t/\mbox{a}\;, \quad f \rightarrow \mbox{a}^2 f\;, \quad h\rightarrow
\mbox{a} \,h\;, \quad C \rightarrow \mbox{a} \,C\;, \quad A_t
\rightarrow \mbox{a} \,A_t\;,
\label{scale1B} \\
&& x \rightarrow x/\mbox{b}\;, \quad B \rightarrow \mbox{b}^2 B\;,
\quad C \rightarrow \mbox{b}\, C\;, \quad A_x \rightarrow \mbox{b}\,
A_x\;,\label{scale2B}\\
&& \label{scale3B}(r, t, x, y, z, L) \rightarrow \alpha
(r, t, x, y, z, L)\;, \quad (A_t, A_x) \rightarrow (A_t, A_x)/\alpha\;,
\\
\label{scale4B}
&& r \rightarrow \beta r\;, \quad (t, x, y, z) \rightarrow (t, x, y,
z)/\beta\;, \quad (A_t, A_x) \rightarrow \beta (A_t, A_x)\;.
\eea
We can forget the third scaling symmetry by setting the length scale $L=1$.  The remaining scalings will be very useful in what follows.

So far our setup is identical to that in \cite{ABKP}. The crucial difference is that the IR boundary conditions we will impose now 
are those of an IR AdS space, not a black hole horizon. In our coordinates, IR corresponds to $r \rightarrow 0$ and UV to 
$r\rightarrow \infty$. Our aim now is to construct a domain wall solution that interpolates between two AdS vacua, corresponding 
to the flow between the symmetry-preserving (UV) and symmetry-breaking (IR) vacua of the scalar potential. To compute the IR expansion, 
we need to know the IR AdS background. The IR AdS scale is fixed by the value of the scalar potential at its symmetry-breaking minimum 
in the action above. The minimum is at
\bea
\psi_0=\operatorname{ArcCosh} 2\,,
\eea
and the effective AdS scale there can be computed from the action to be
\beq
L_{IR}=\frac{2^{3/2}}{3} L\,.
\eeq
Unlike the finite temperature case, the correct IR boundary expansions of the fields are now no longer of a simple power series form in $r$. The easiest way to see this is to note that the scalar field equation in the IR AdS background allows the solution
\beq
\psi = 
 \operatorname{ArcCosh} 2 + \psi_{1,0} \ r^{\alpha}\,,
\eeq
with $\psi_{1,0}$ an integration constant. The reason for choosing this specific form for the subscripts on this coefficient will become clear momentarily. Here, $\alpha$ can be computed\footnote{This value was computed also in \cite{GubRoch}, but the value reported there has a typo.} by solving the scalar field equation in the IR AdS, and the result is
\beq
\alpha=2 \sqrt{3}-2 \simeq 1.46\, .
\eeq
Since we are looking for a non-trivial solution that starts off at the symmetry-breaking  minimum for small $r$, we should allow for such irrational powers in the IR expansion when trying to extrapolate from the IR to UV. We will try out an expansion\footnote{We thank S. Gubser and S. Pufu for helpful discussions on this.} for the fields by writing them in powers of the two variables $x$ and $y$ defined as
\beq
x\equiv r^\alpha\,, \qquad y \equiv r\,,
\eeq
and then constrain the coefficients by demanding that they satisfy the equations of motion.  This fixes the form of the fields in the IR (up to ${\cal O}(r^2)$) to be
\bea
&&f(r)=\frac{9}{8}h_{0,0}^2+\frac{4\ A_{t(0,2)}^2}{3}\ y^2+...\;,\qquad
h(r)=h_{0,0}+h_{0,2}\ y^2+...\; , \label{h00}\\
&&B(r)=B_{0,0}-\frac{4 A_{x(0,2)}^2}{3}\ y^2+...\;, \qquad\; C(r)=C_{0,0}+\frac{4 A_{x(0,2)}A_{t(0,2)}}{3}\ y^2+...\;, \\
&&A_t(r)=A_{t(0,2)}\ y^2+...\;, \qquad\qquad\quad\;\;  A_x(r)=A_{x(0,2)}\  y^2+...\;, \\
&&\psi(r)=\operatorname{ArcCosh} 2 + \psi_{1,0} \ x+\psi_{0,2} \ y^2+ ... \;,
\eea
where not all of the coefficients are independent. In particular
\bea
&&h_{0,2}=\frac{2  (-27 h_{0,0}^4\, A_{x(0,2)}^2 + 64 C_{0,0}^2\, A_{t(0,2)}^2 + 
   48 h_{0,0}^2\, C_{0,0}\, A_{t(0,2)}\, A_{x(0,2)}  + 96 h_{0,0}^2\,B_{0,0}\,A_{t(0,2)}^2)}{27 (8 C_{0,0}^2\,h_{0,0} + 9 B_{0,0} \,h_{0,0}^3)}\;, \nonumber \\
\\
&&\psi_{0,2}=\frac{16  (9 h_{0,0}^2\, A_{x(0,2)}^2 - 
     16 C_{0,0}\, A_{t(0,2)}\, A_{x(0,2)}- 8  B_{0,0}\, A_{t(0,2)}^2)}{3 \sqrt{
   3} (8 C_{0,0}^2 + 9 B_{0,0}\, h_{0,0}^2)}\,. \hspace{1in}
\eea
The notation for the coefficients should be clear by now: in the IR expansion for a field $\phi$, the coefficient of $x^a y^b$ is denoted as $\phi_{a,b}$.
All higher order coefficients (which include powers of $x$, powers of $y$ and mixed powers) are fixed iteratively in terms of the lower order coefficients due to the equations of motion. We will need the IR expansion to start the numerical integration from the IR, but the explicit forms of the higher coefficients are not very enlightening, so we will not report them here. Note, finally, that there are only six independent quantities in the IR, just as there were at the horizon in \cite{ABKP}. In the present case, these independent quantities are
\beq
\{h_{0,0}\,,\; B_{0,0}\,,\; C_{0,0}\,,\; A_{t(0,2)}\,,\; A_{x(0,2)}\,,\; \psi_{1,0} \}\,.
\eeq
Our strategy in constructing the solutions will be to pick numerical values for these IR data, and integrate the equations of motion numerically all the way to some very large $r$, corresponding to the UV. At the UV boundary, we can again solve the equations of motion straightforwardly and look for a consistent expansion that reproduces the curves arising from this integration. The relevant series expansions can be found in eqns. (3.18)-(3.23) and Appendix A of  \cite{ABKP} and we will use the notations there in our UV discussion. To get asymptotically AdS boundary conditions in the UV, we need to set $B_0=1=h_0$, and $C_0=0=\psi_1$. The former two conditions can be accomplished via the rescalings (\ref{scale1B}, \ref{scale2B}), while a shooting technique is required for the latter two. This gives rise to the eight independent boundary quantities \cite{ABKP}:
\beq
(f_4\,,\; B_4\,,\; C_4\,,\; A_{t,0}\,,\; A_{t,2}\,,\; A_{x,0}\,,\; A_{x,2}\,,\; \psi_3)\,. 
\eeq
We will also use the rescaling (\ref{scale4B}) to set the leading piece of $A_t$ at the boundary (namely $A_{t,0}\equiv \mu$) to unity\footnote{In \cite{GubRoch} this scaling was used to set the horizon datum $A_{t(0,2)}=1$. We prefer instead to set $A_{t,0}=1$ at the boundary: this corresponds to working in a fixed chemical potential ensemble in the gauge theory.}. Note that since this rescaling involves $r$, we will need to appropriately rescale the coefficients of the IR expansion as well, when launching the integration\footnote{In the finite temperature case of \cite{ABKP}, this scaling shifts the horizon radius and therefore effectively introduces a new parameter, the temperature.}. Once we fix the chemical potential to one, for any given superfluid velocity $\xi$ (before the rescaling of the solution that sets $A_{t,0}$ to unity, $\xi$ is given by $A_{x,0}/A_{t,0}$)  the number of independent parameters at the boundary is six, which is the same as the number of horizon data. Therefore we expect to find at most discretely many domain wall solutions for any given superfluid velocity. As discussed in \cite{GubRoch} we will look for the solution where the radial profile of the scalar field has the least number of nodes.

\section{Results}

After the shooting and the rescaling discussed in the previous section, we have the domain wall solution we were seeking. We present the plots of the various functions for a selected value of the velocity in Figure 2.
\begin{figure} [ht]
\begin{center}
\includegraphics[width=\textwidth]{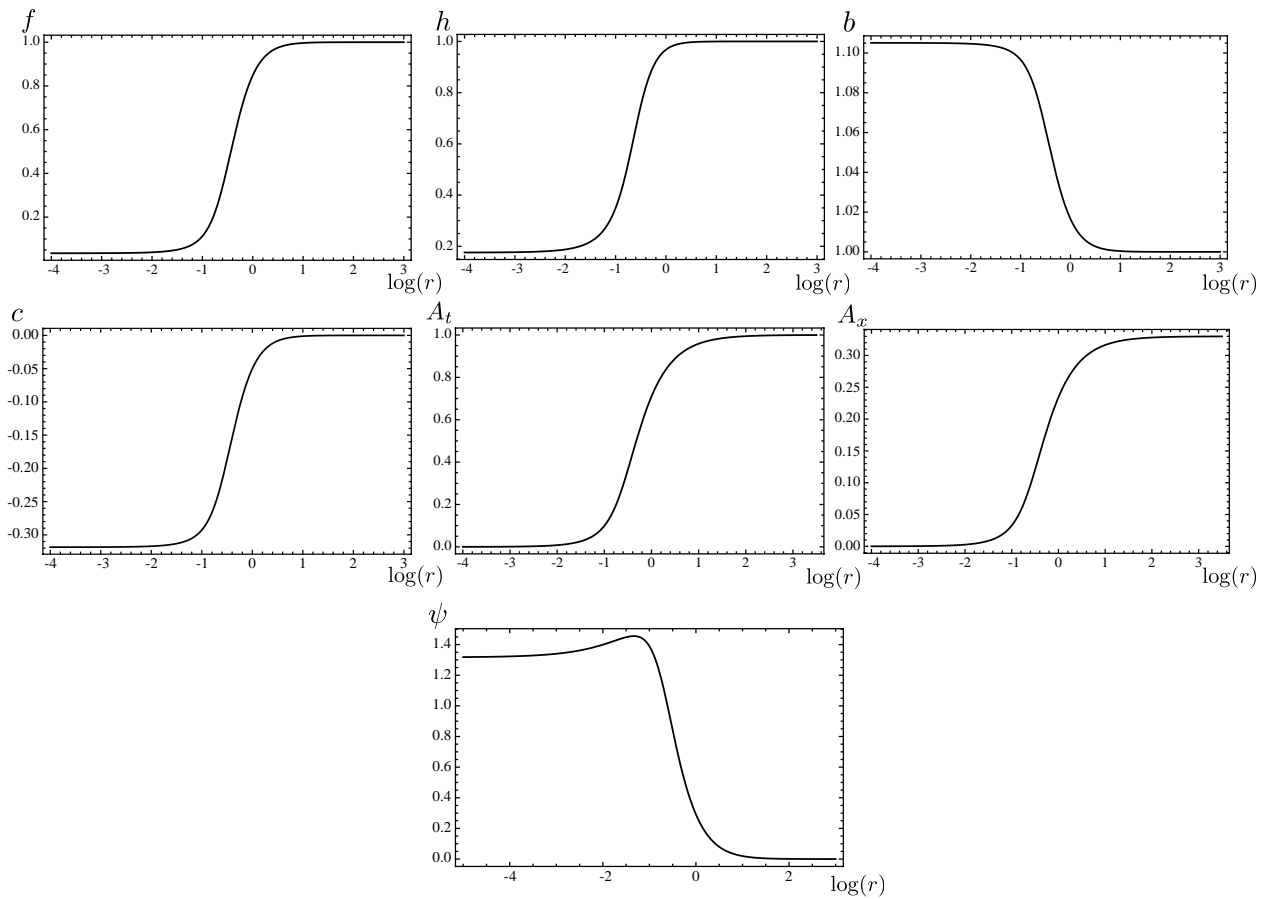}\label{dwalls}
\caption{Plots of the various functions as a function of the radial coordinate for $\xi=0.33$. A logarithmic radial coordinate has been chosen to illustrate that the solution is a domain wall. For other values of the superfluid velocity in the range $0<\xi<\xi_c$ the nature of the curves is similar.}
\end{center}
\end{figure}
We have checked that the plots for other velocities are qualitatively similar.
We find that solutions exist for velocities $\xi$ ranging from $0$ to $\xi_c \approx 0.374$, which is consistent with the expectation of \cite{ABKP}: there we found evidence that an AdS IR fixed point exists for low enough velocities but not for $\xi \ge 0.40$, even though the finite-temperature superfluid phase exists to a higher value $\xi_*\sim 0.5$. It was also found there that the condensate, as defined by
\beq
\langle O_{DW}\rangle =\frac{(\sqrt{2}\psi_3)^{1/3}}{\sqrt{1-A_{x,0}^2}}
\eeq
tends to a fixed value as the temperature is lowered, for small enough velocities. (In writing the above expression for the condensate, we use the fact that $A_{t,0}$ has been rescaled to one.) We can compute the same quantity using our domain wall solutions, and we find that for all the velocities for which it exists, the value of this condensate is a constant and is precisely equal to the one found as the zero-temperature limit of our finite temperature solutions. We can also perform another check of our solutions, by considering the isotropic (i.e., $\xi=0$) limit. In 
that case, it was found in \cite{GubRoch} that the object $\langle O \rangle_{GPR} \equiv \frac{\psi_3}{(2/\sqrt{3})^3} \approx 0.322$. 
We have checked that 
our solutions reproduce this value in the isotropic limit. 

The solutions we constructed were obtained via a shooting method. So one might worry that the absence of solutions above $\xi\approx0.374$ is an artifact of the numerics. It is difficult to disprove this conclusively, but we can look at the profiles of some quantities as the velocity is changed and see what happens to them as the critical velocity $\xi_c$ is approached. One such useful quantity is the refractive index between the UV and the IR. We can define this quantity as the ratio of the propagation velocities of light in the UV and IR
\beq
n=\sqrt{\frac{f_{UV}}{f_{IR}}}
\eeq
where $f$ is contained in the time component of the metric (\ref{ModTisza4}). We plot its behavior as a function of the superfluid velocity in Figure 3. 
\begin{figure} [ht]
\begin{center}
\includegraphics[height=0.3\textheight]{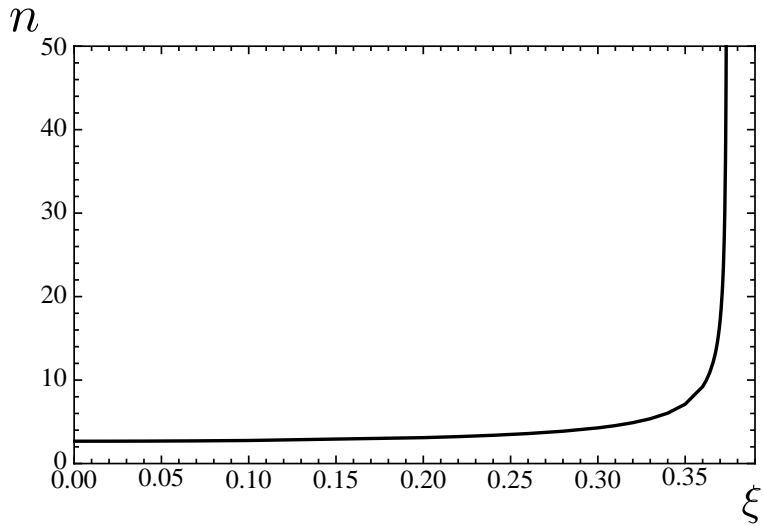}\label{refrc}
\caption{Plot of the refractive index as a function of the superfluid velocity. The refractive index diverges around $\xi_c\approx 0.374$, where the IR AdS space collapses and the domain walls we have constructed cease to exist. In the isotropic limit the numerical value of the refractive index is $n \approx 2.674$ which is identical to what is reported in \cite{GubRoch2}.}
\end{center}
\end{figure}
It is clear from the figure that at $\xi_c$ it diverges, strongly suggesting that domain wall solutions of this kind do not exist beyond $\xi_c$. It turns out that the AdS space in the IR degenerates as $\xi \rightarrow \xi_c$ because $h_{0,0}$ in the IR expansion (\ref{h00}) tends to zero.

\section{Interpretation and discussion}

Let us summarize our results and discuss them in light of the Criticality Pairing Conjecture  (CPC) proposed in  \cite{GubRoch}. We can state 
the CPC as follows. Consider a theory with a $U(1)$ symmetry that, once deformed by some appropriate operator, flows to an IR fixed point that breaks the $U(1)$. Then the claim is that the theory or its deformations by operators that do not break the $U(1)$ explicitly has a finite density, zero temperature state whose infrared behavior is controlled by the same IR fixed point. 

The deformation considered in \cite{GubRoch} is given by $\mu J^0$. This operator doesn't break the $U(1)$ symmetry, and hence satisfies the premise of the CPC. The explicit construction of the domain wall in \cite{GubRoch} demonstrates that the deformation by this operator results in a flow to AdS in the IR. In this paper, we considered a more general deformation, $\mu J^0+A_{x,0} J^x$, which breaks isotropy but which again does not break the $U(1)$ symmetry. We find that when this more general deformation is added, the system is controlled by the same AdS IR fixed point, in agreement with the CPC. However, this holds only as long as the source term, $A_{x,0}$,  is small enough: for high enough velocity we seem to find a counter-example to the CPC. 

A violation of the CPC was presented in \cite{Nellore} where for small enough values of the charge of the scalar the IR fixed point was destabilized. In our case, the parameter is not directly visible in the supergravity Lagrangian, and therefore the destabilization that happens at large velocities is of a different kind. Note that since we are working at zero temperature and the theory is conformal, all non-zero values of the chemical potential are equivalent. But when we turn on $A_{x,0}$ as well, we have a tunable parameter, namely the superfluid velocity $\xi=A_{x,0}/\mu$. Once this becomes too large, the anisotropy becomes too strong to be washed out in the IR. This is similar to what happens in ordinary superconductors: when the chemical potential imbalance becomes large, they enter an anisotropic FFLO phase \cite{comb,FFLO}. 

In fact, one of the questions we have left unanswered in this paper is that about the possibility of constructing a useful zero-temperature limit for the superfluid flows in the range $\xi_c< \xi < \xi_*$. As already mentioned, in \cite{Nellore} it was found that the CPC can be violated if the scalar charge is tuned to be small, and in such case a Lifshitz geometry 
emerges in the infrared. This is not unreasonable because spatial isotropy is not broken in that case - only a charge density is turned on, and no current. In our case, spatial isotropy is broken\footnote{As we have seen, this is not a sufficient condition for the IR geometry to be anisotropic, but it is a hint. } and our preliminary investigations suggest that the system is not controlled by a simple Lifshitz (or generalized Lifshitz \cite{Kiritsis}) geometry. This can be understood as follows.  A geometry of the form
\beq
ds^2=-r^{z_t}dt^2+{r^{z_r}}{dr^2}+r^{z_x}dx^2+r^{z_y}dy^2+r^{z_z}dz^2
\eeq
has curvature scalars $R$ and $R_{abcd}R^{abcd}$ that behave as 
\beq \label{constr}
R \sim \frac{1}{r^{z_r+2}}\;,\qquad R_{abcd}R^{abcd} \sim \frac{1}{r^{2z_r+4}} \,.
\eeq
(Generalized) Lifshitz corresponds to the case $z_r=-2$ \cite{Kiritsis}, and then both quantities are constants. But numerically, as we lower the temperature of the finite temperature solutions, we find that there is no region close to the horizon where both $R$ and $R_{abcd}R^{abcd}$ stabilize simultaneously. This could of course be a limitation of the numerics, but together with the fact that this is happening at high velocities is a suggestion that if an IR geometry exists, its anisotropy should take a different form. 
Allowing for  $z_r \not =-2$ might be another possibility\footnote{But our preliminary numerical results are consistent with $L^2 h^2/f$ (see \ref{ModTisza4}) stabilizing to a constant as the temperature is lowered, so this might not be the case.}, so is the possibility of allowing non-trivial scalar and gauge field profiles in the IR. Note, however, that just allowing non-trivial matter fields will not bypass the constraints on the curvatures (\ref{constr}).
The bottom line is that it would be interesting to construct such an anisotropic AdS-to-nonAdS domain  wall, if it exists. Our preliminary attempts to look for one have been inconclusive. As reported in \cite{ABKP}, at finite but small temperature the curvature scalars begin to grow quite quickly as we approach the horizon, so it is not obvious that there is a sensible way in which one can assign an IR geometry to this flow. 
It is possible that the natural interpretation is that for velocities greater than $\xi_c$ the superfluid exhibits runaway in the IR and not a fixed point. Answering this question 
might also let one understand to what extent the analogy between holographic superfluids at finite superfluid velocity and superconductors with imbalanced chemical potential, 
can actually be pushed.

\section*{Acknowledgments}

We would like to thank I. Amado, A. Cotrone, J. Gauntlett, S. Hartnoll, E. Kiritsis, J. Sonner and especially S. Gubser and S. Pufu 
for discussions/correspondence.  D.~A. thanks SISSA for hospitality and the FRont Of Galician-speaking Scientists for unconditional 
support. C.~K. and T.~P. thank ICTP for hospitality.


\end{document}